\documentclass[journal=ancac3,manuscript=article]{achemso}

\usepackage{color}
\usepackage{multirow}
\usepackage{epsfig}
\usepackage{graphicx}
\usepackage{amsmath}
\usepackage{amssymb}
\usepackage{bm}
\usepackage{dcolumn}
\usepackage{braket}
\usepackage{bbold}
\usepackage{array}
\usepackage{tabulary}
\usepackage{float}
\usepackage{CJKutf8}
\usepackage{lipsum}
\usepackage[normalem]{ulem}
\usepackage{xkeyval,xcolor}

\title{Weyl nodal point-line Fermion in ferromagnetic Eu$_5$Bi$_3$}

\author{Hongbo Wu$^{1}$}
\author{Da-Shuai Ma$^{1}$}
\author{Botao Fu$^{2}$}
\email{fubotao2008@gmail.com}
\author{Wei Guo$^{1}$}
\author{Yugui Yao$^{1}$}
\email{ygyao@bit.edu.cn}

\affiliation{$^1$
    Key Laboratory of advanced optoelectronic quantum architecture and measurement (MOE), Beijing Key Laboratory of Nanophotonics and Ultrafine Optoelectronic Systems, School of Physics, Beijing Institute of Technology, Beijing 100081, China \\
	$^2$ College of Physics and Electronic Engineering, Center for Computational Sciences, Sichuan Normal University, Chengdu, 610068, China}
\date{\today}

\begin{document}

\begin{abstract}
	Based on $ab$ $initio$ calculations and low-energy effective $\bm{k}\cdot\bm{p}$ model, we propose a type of Weyl nodal point-line fermion, composed of 0D Weyl points and 1D Weyl nodal line, in ferromagnetic material Eu$_5$Bi$_3$. In the absence of spin-orbital coupling (SOC), the spin-up bands host a pair of triply degenerate points together with a unique bird-cage like node structure. In the presence of SOC with (001) magnetization, each triplet point splits into a double Weyl point and a single Weyl point accompanied by two nodal rings, forming two sets of Weyl nodal point-line fermions near the Fermi level. The novel properties of Weyl nodal point-line fermion are explored by revealing the unusual Berry curvature field and demonstrating the pinned chiral surface states with exotic Fermi arcs at different planes. Moreover, a large anomalous Hall conductivity of -260 ($\hbar$/$e$)($\Omega$cm)$^{-1}$ parallel to [001] direction is predicted. Our work offers a new perspective for exploring novel topological semimetal states with diverse band-crossing dimensions, and provides an ideal material candidate for future experimental realiztion.

\end{abstract}

\maketitle

Topological semimetals (TSMs), emerging as new types of topological states, have recently attracted plenty of interest due to their physical meanings as counterparts for fundamental particles in field theory such as Weyl fermion and Dirac fermion~\cite{armitage2018weyl,young2012dirac,wang2012dirac,xu2015discovery,chen2015nl,weng2016topology,chiu2016classification,huang2018nl,HirayamaMtop2018,jiang2019nl} and also due to their fascinating properties like chiral anomaly~\cite{zyuzin2012topological,son2013chiral,gorbar2014chiral,huang2015observation}, surface magnetism\verb|/|superconductivity\cite{heikkila2011flat,qi2016superconductivity,chan20163,wang2017topological} and 3D quantum hall effect (QHE)\cite{brune2011quantum,wang20173d,zhang2018quantum}, etc. The band structures of TSM materials possess nontrivial band crossings close to the Fermi level and can be characterized by various topological invariants. Depending on the dimensionality of the band crossings, TSM states can be classified into zero-dimensional (0D) nodal point\cite{burkov2011weyl,lv2015experimental,soluyanov2015type}, one-dimensional (1D) nodal line\cite{burkov2011topological,mullen2015line,fang2016topological,hu2016evidence,li2017type,fang2015topological} and two-dimensional (2D) nodal surface\cite{zhong2016towards,liang2016node,bzduvsek2017robust} semimetals. The band crossings with distinctive dimensions lead to various topology and interesting physical properties.

Under broken time reversal ($T$) or space inversion ($P$) symmetry, Weyl semimetals (WSMs) that host twofold band-crossing points at the Fermi level can be achieved in magnetic or noncentrosymmetric materials\cite{weng2015weyl,yan2017topological,burkov2018weyl,wang2018nodal,fengbaojie2019}. Weyl point (WP) and Weyl nodal line (WNL) semimetals are currently the most prominent compositions of WSMs\cite{lv2015experimental,soluyanov2015type,fang2016topological}. The topological nature of 0D WPs can be characterized by the non-zero chern number and exotic surface Fermi arcs\cite{wan2011topological,hasan2017discovery,belopolski2016fermi,deng2016experimental}.
When an additional rotation symmetry exists (e.g. $C_3$, $C_4$, $C_6$), multiple WPs will merge together at the rotation-invariant axis to create a high chern number Weyl fermion, which exhibits parabolic or cubic dispersion\cite{fang2012multi,jian2015correlated,huang2016new,Tsirkin2017Composite,zaheer2013spin,weng2016topological,winkler2016topological,zhu2016triple,yang2017prediction,gao2018possible,zhang2017jpcl}.
Unlike the WP semimetal, a WNL semimetal possesses a closed loop of double degenerate points in the momentum space and can be diagnosed by a quantized Berry phase through SdH oscillation experiment\cite{hu2016evidence} or a flat "drumhead"-like surface state by ARPES\cite{neupane2016observation,bian2016topological,wang2016evidence}. A WNL structure is generally protected by a mirror or slide mirror symmetry\cite{gao2016classification,yang2017symmetry} under strong SOC effect. The individual WP semimetal and WNL semimetal have been widely studied and part of them have already been observed in recent experiments\cite{xu2015discovery,lv2015experimental,soluyanov2015type,fengbaojie2019,deng2016experimental,bian2016topological,kim2018large,wang2018large}. Motivated by present fascinating topological harvest, we ask a question that can those topological semimetal states with diverse dimensions coexist in one system and what novel properties it will bring about if it exists.

In this paper, based on first-principles calculations and effective $\bm{k}\cdot\bm{p}$ model, we propose a concept of Weyl nodal point-line (WNPL) fermion  in a ferromagnetic material Eu$_5$Bi$_3$, which consists of a pair of 0D WPs and a 1D WNL that couple together in one physics system and can be created and annihilated simultaneously via band inversion mechanism.  In the case without SOC , the spin-polarized bands of Eu$_5$Bi$_3$ host a pair of triple points in HKH$^{\prime}$ path protected by rotation symmetry. Contrast to other known triplet points in nonmagnetic and non-centrosymmetric materials\cite{zaheer2013spin,weng2016topological,winkler2016topological,zhu2016triple,yang2017prediction,gao2018possible}, this two magnetic triplet points connect to each other by an unique bird-cage like node structure centering at K point. When SOC effect is included and consider (001) magnetization, a topological phase transition happens. Each triplet point splits into a double Weyl point and a single Weyl point accompanied by two homocentric WNLs at $k_z=0$ plane, thus forming two sets of composite WNPL fermions around the Fermi level. Moreover, we build a low-energy effective $\bm{k}\cdot\bm{p}$  model that can capture the novel topological nature of the WNPL fermion. In particular, we discover that the WNPL fermion possesses strong coupling behavior that  gives rise to pinned chiral surface states with exotic Fermi arcs at different projection planes.

\begin{figure}[h]
	\includegraphics[width=7.2cm]{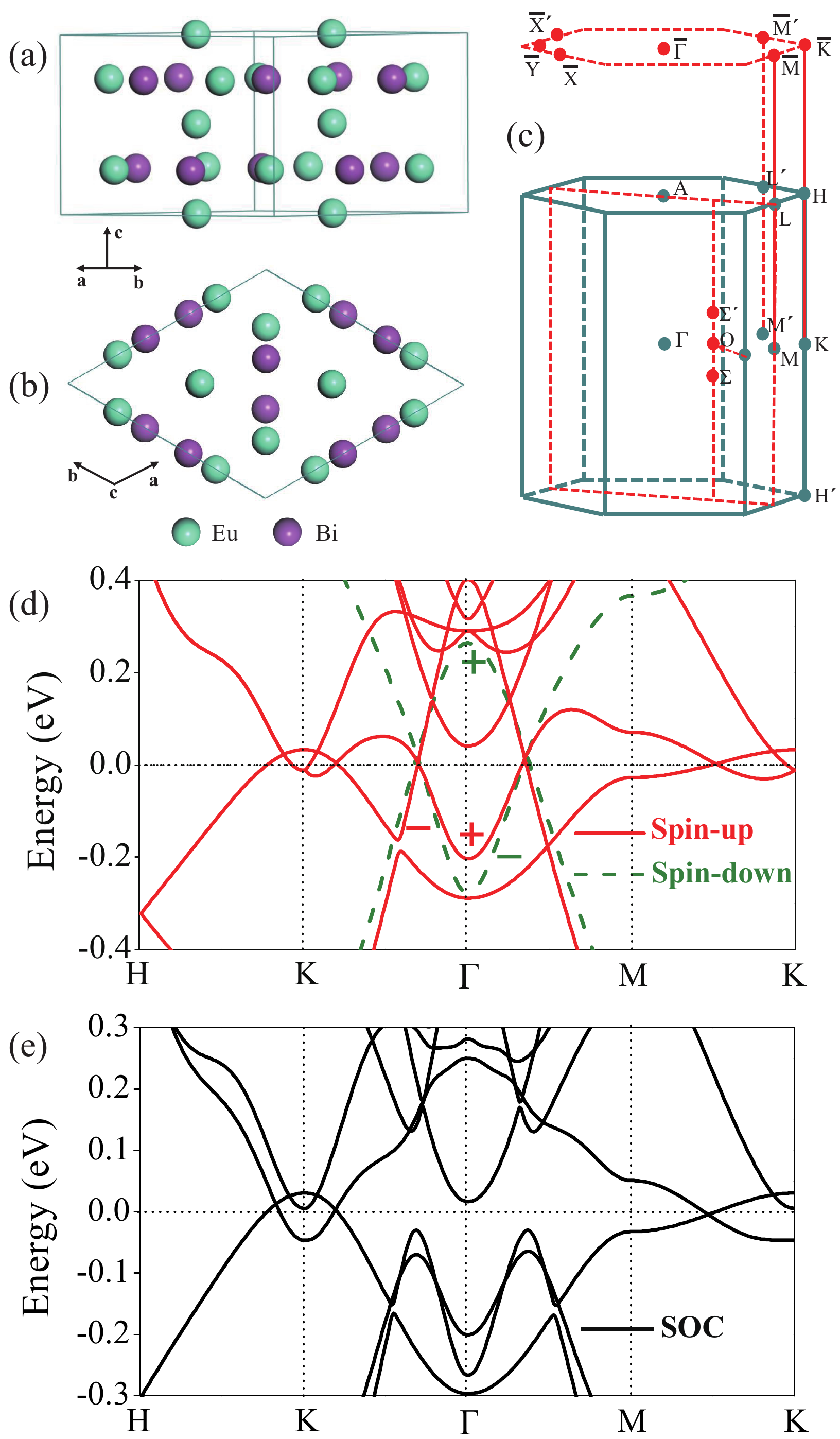}
	\caption{Crystal structure, Brillouin zone and band structure of Eu$_5$Bi$_3$. (a) Side view and (b) top view of the crystal structure of Eu$_5$Bi$_3$ with space group $P6_3/mcm$ (No. 193). (c) The hexagonal Brillouin zone, the high-symmetry $\bm{k}$ path and the projected surface Brillouin zone of (001) and (100) planes. The red dashed lines denote the projected surface Brillouin. (d) Spin-polarized band structure of Eu$_5$Bi$_3$. (e) SOC band structure of Eu$_5$Bi$_3$ with (001) magnetization. The Fermi level is shifted to zero in (d) and (e).}\label{FIG.1}
\end{figure}

The rare-earth bismuthide Eu$_5$Bi$_3$ crystallizes in a stacked hexagonal lattice with the space group $P6_3\verb|/|mcm$ (No. 193). It hosts a point group $D_{6h}$ that includes a six-fold rotation axis $C_{6z}$, a horizontal mirror $\sigma_h$ and six vertical mirror $\sigma_v$. The geometric structures of Eu$_5$Bi$_3$ are shown in Figs. \ref{FIG.1} (a)-(b), where Eu atoms take two types of Wyckoff positions 4$c$ (1\verb|/|3, 2\verb|/|3, 0) and 6$g$ (0.251, 0, 1\verb|/|4), and Bi atoms take one type of Wyckoff position 6$g$ (0.609, 0, 1\verb|/|4). Meanwhile, the hexagonal bulk Brillouin zone with high-symmetry $\bm{k}$ paths and projected (001)\verb|/|(100) surface Brillouin zone are illustrated in Fig. \ref{FIG.1}(c).
The optimized lattice constants of Eu$_5$Bi$_3$ are $a$ = 9.68 \AA {} and $c$ = 7.36 \AA, in good agreement with the experimental data\cite{leon2006hydrogen}.
The macroscopical ferromagnetic states of Eu$_5$Bi$_3$ had been observed below 101 K\cite{leon2006hydrogen}. Our first-principles calculations\cite{supplement} also confirm that the ferromagnetic (FM) state is the ground state with lowest total energy and each Eu atom possesses a magnetic moment of $\sim$ 7.0 $\mu_{B}$, which is in good accordance with the experimental value\cite{leon2006hydrogen}. In addition, the calculated magnetic crystalline anisotropy energy is less than 0.05 meV\verb|/|atom between out-plane and in-plane magnetization, indicating the switch of magnetization direction by an external magnetic field is operable.

The spin-polarized band structure of Eu$_5$Bi$_3$ without SOC is demonstrated in Fig. \ref{FIG.1} (d).
In the absence of SOC, spin is a good quantum number.
We notice that both spin-up and spin-down channels exhibit topological metal states with various band crossings near the Fermi level. For the spin-down bands (green dashed line) at low-energy part, the conduction band crosses with the valence band forming a clean and large-size nodal line structure encircling $\Gamma$ point. It is easy to confirm that this nodal line structure, located at $k_z = 0$ plane, is protected by the horizontal mirror $\sigma_h$ because of its opposite eigenvalues. Similarly, for the spin-up bands (red solid line), a nodal line structure lying on $k_z = 0$ plane emerges due to the band inversion around $\Gamma$ point.

As both the Eu and Bi atoms belong to heavy elements that have non-negligible SOC effect, we then switch on the SOC and consider (001) magnetization. The band structure with SOC is displayed in Fig. \ref{FIG.1} (e). The point group of Eu$_5$Bi$_3$ transforms from $D_{6h}$ into C$_{6h}$, where the vertical mirror $\sigma_v$ is broken but the horizonal mirror $\sigma_h$ is left. Due to the SOC effect, the spin $s_z$ is no longer a good quantum number. Thus, the spin-up and spin-down bands with same eigenvalue of $\sigma_h$ that cross each other at the Fermi level in Fig. \ref{FIG.1} (d) will inevitably couple each other, and a gap is opened around $\Gamma$ point as shown in Fig. \ref{FIG.1} (e).

\begin{figure}[h]
	\includegraphics[width=9cm]{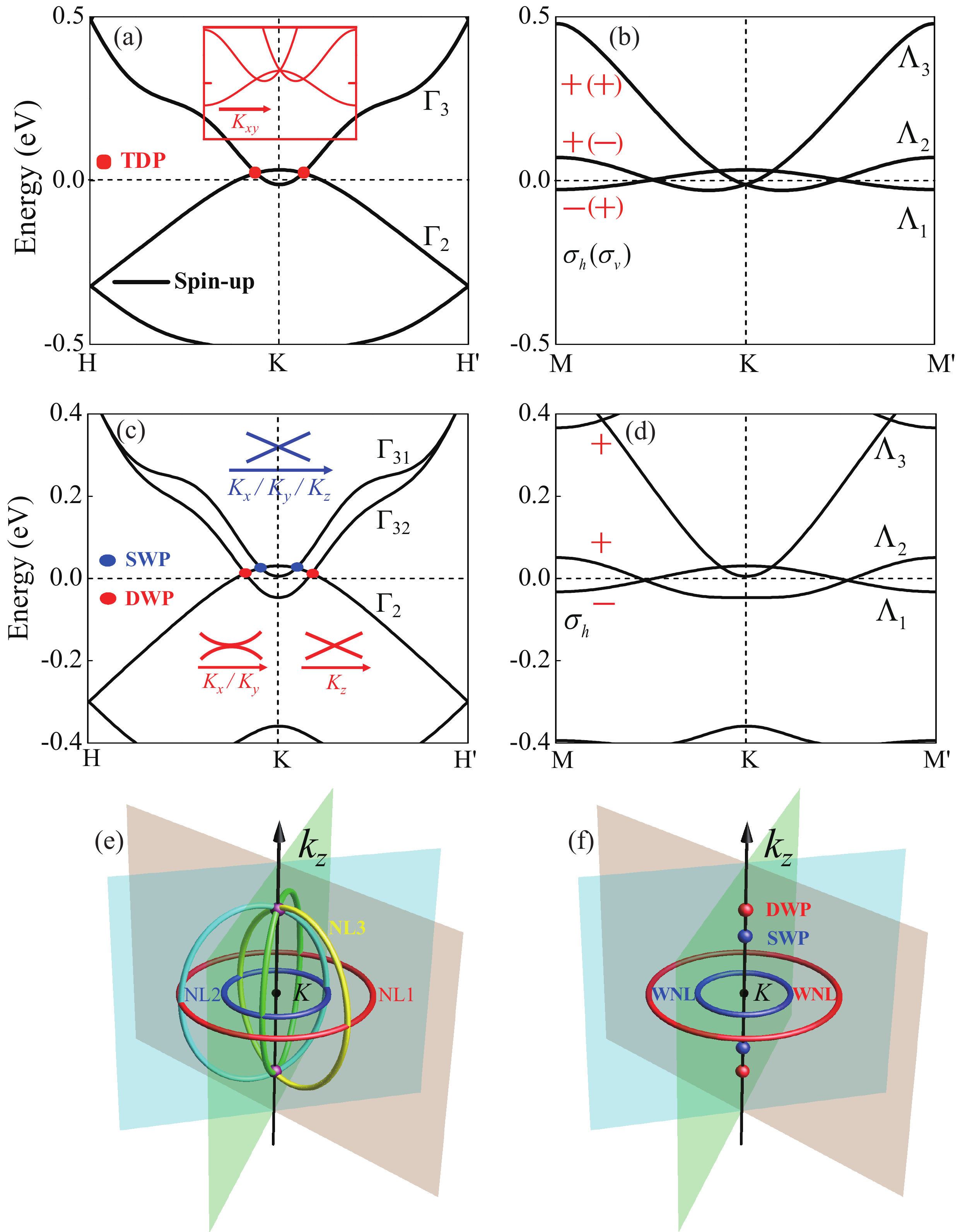}
	\caption{Band structures of Eu$_5$Bi$_3$ focusing on K point. Spin-up/SOC band structures of Eu$_5$Bi$_3$ along high-symmetry (a)/(c) H-K-H$^{\prime}$ and (b)/(d) M-K-M$^{\prime}$ paths. $K_{xy}$ direction is parallel to $k_{x}+k_{y}$ direction. Mirror eigenvalues of $\sigma_h$ and $\sigma_v$ are labeled in  figure (b) and (d). (e) and (f) show the 3D view of the band crossing points around K point with and without SOC, respectively. The Fermi level is shifted to zero in (a), (b), (c) and (d).}\label{FIG.2}
\end{figure}

In this section, we focus on the band structures of Eu$_5$Bi$_3$ around K point. In the absence of SOC, the little group along H-K-H$^{\prime}$ path is $C_{3v}$, which hosts a two-dimensional and a one-dimensional irreducible representations that can support a two-fold and a singlet band degeneracy, respectively. Consequently, as shown in Fig. \ref{FIG.2} (a), a doubly degenerate conduction band ($\Gamma_3$) inverts with a singlet valence band ($\Gamma_2$) and forms a pair of triplet degenerate points (TDPs) along H-K-H$^{\prime}$ path.
The negative band gap at K point from the band inversion is sensitive to Hubbard U due to strongly correlated $f$ electrons in Eu atoms, nevertheless, we have confirmed the band inversion does happen within a wide and reasonable on site Coulomb interaction (see the Supporting Information Fig.S 1)\cite{supplement}.
With inset in Fig. \ref{FIG.2} (a), we demonstrate the dispersion of the TDP in $K_{xy}$ direction, and find a linear band crossing superimposed with a quadratic band which is similar to that in non-magnetic TDP systems\cite{zhu2016triple}.

As shown in Figs. \ref{FIG.1} (c), the M-K-M$^{\prime}$ path locates at the intersection of $\sigma_h$ invariant plane (${\Gamma}$MKM$^{\prime}$) and $\sigma_v$ invariant plane (MKHL).
Thus, the energy bands along M-K-M$^{\prime}$ can be represented by the eigenvalues of $\sigma_h$ and $\sigma_v$, respectively as shown in Fig. \ref{FIG.2} (b). We discover two conduction bands ($\Lambda_{2}$, $\Lambda_{3}$) possess positive eigenvalue of $\sigma_h$, while the valence band ($\Lambda_1$) has negative eigenvalue. As a result, at the $\sigma_h$ invariant plane ${\Gamma}$MKM$^{\prime}$ , the crossing points of bands $\Lambda_{2}$ and $\Lambda_{1}$ have to form a nodal line structure (NL1 in Fig. \ref{FIG.2} (e)) centered at K point. Similarly, the crossing points between bands $\Lambda_{3}$ and $\Lambda_{1}$ also form a nodal line structure (NL2). The NL1 structure encompasses NL2 at $k_z$=0 plane, and NL1 is more closer to Fermi level than NL2.
The eigenvalues of $\sigma_v$ are also calculated as shown in Fig. \ref{FIG.2} (b). We find two conduction bands $\Lambda_{2}$ and $\Lambda_{3}$ possess negative and positive eigenvalues, respectively, and the conduction band $\Lambda_{1}$ has positive eigenvalue. Thus, at $\sigma_v$ invariant plane MKHL, the band crossings of $\Lambda_{2}$ and $\Lambda_{1}$ are mirror symmetry protected and form a nodal line structure (NL3). On the contrary, the band crossings between band $\Lambda_{3}$ and $\Lambda_{1}$ are avoidable because of the same eigenvalues of $\sigma_v$. Considering the $C_{3z}$ rotation of HKH$^{\prime}$ axis, three equivalent NL3 emerge and connect to each other exactly at the TDPs. As schematically shown in Fig. \ref{FIG.2} (e), the NL3 structure at extended MKHL plane is not eudipleural with respect to HKH$^{\prime}$ axis and noted as an off-centered nodal ring because of polar $C_{3z}$ symmetry. In addition, the NL1 and NL2 at $k_z=0$ plane interconnect with three vertical NL3 from inside and outside, respectively, thus forming a cage-like node structure encircling K point. This unique nodal line interconnection makes this ferromagnetic triplet fermion distinguished from other known TDP semimetal states\cite{weng2016topological}.

When the SOC effect is included with (001) magnetization, the little group along H-K-H$^{\prime}$ is lowered to $C_3$ that can not guarantee a two-dimensional representation any more. As a consequence, the doubly degenerate conduction band ${\Gamma}_3$ in Fig. \ref{FIG.2} (a) splits into two singlet states ($\Gamma_{31}$ and $\Gamma_{32}$) in Fig. \ref{FIG.2} (c). Accordingly, a pair of TDPs are splitting into two pairs of WPs along H-K-H$^{\prime}$ path.
By calculating the charities of the WPs, we identify that the WPs between $\Gamma_{32}$ and $\Gamma_{2}$ are double Weyl points (DWPs, C = $\pm$ 2) while the WPs between $\Gamma_{31}$ and $\Gamma_2$ are single Weyl points (SWPs, C = $\pm$ 1).
Remarkably, the strong SOC effect in ferromagnetic Eu$_5$Bi$_3$ transforms a triplet point into two WPs with different charities.
We have identified the distinctive charities stem from distinctive eigenvalues of $C_{3}$ operation for $\Gamma_{31}$ and $\Gamma_{32}$\cite{supplement}.
On the other hand, the SOC breaks $\sigma_h$ symmetry but reserves ${\sigma}_v$ symmetry. Therefore, the NL3 is broken but NL1 and NL2 are reserved. As schematically demonstrated in Fig. \ref{FIG.2} (f), the Eu$_5$Bi$_3$ system simultaneously host 0D WPs and 1D WNLs coupled around K point, thus forming a new type of WNPL Fermion around the Fermi level. This composite fermion is quite different from other's proposals of a simple coexistence of WPs and WNL without physical relevance\cite{Rauch2017Model,Sun2017Coexistence}. The composite WNPL fermion in Eu$_5$Bi$_3$ is derived from band inversion and protected by $C_{6v}$ point group. Our low-energy effective $\bm{k} \cdot \bm{p}$ model will further reveal the intrinsic coupling nature of the WNPL fermion.


\begin{figure}[h]
	\includegraphics[width=6.8cm]{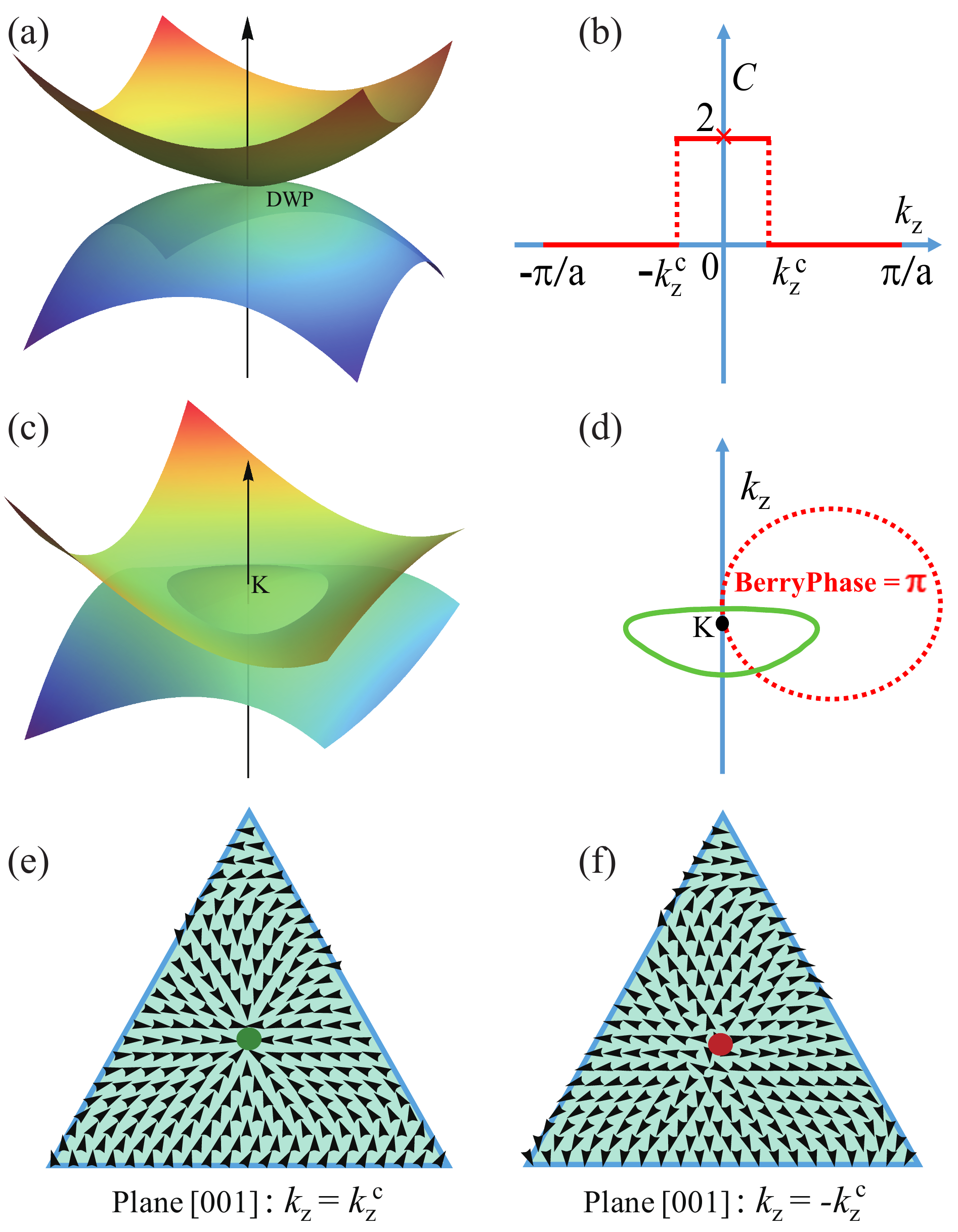}
	\caption{3D view of the band dispersions of (a) DWP and (c) WNL. (b) Schematic of the Chern number as a function of $k_z$ along H-K-H$^{\prime}$ direction. (d) Quantized Berry phase of a arbitrary loop intersect a target WNL with magnetization parallel to [001] direction. (e) and (f) show the distribution of the Berry curvature of DWP with negative and positive chirality at $k_z$ = $k_{z}^{c}$ and $k_z$ = -$k_{z}^{c}$ plane, respectively.}\label{FIG.3}
\end{figure}

In fact, the ferromagnetic Eu$_5$Bi$_3$ host two sets of WNPL structures centred at K point. As schematically plotting in Fig. \ref{FIG.2} (f), one is composed of a pair of DWPs and a WNL structure (red color), and the other is consisted of a pair of SWPs and a WNL structure (blue color).
Here, taking the first WNPL structure (closer to Fermi level) for example, the corresponding two-band effective $\bm{k} \cdot \bm{p}$ model can be written as\cite{supplement}
\begin{eqnarray}
\mathcal{H}\left(\boldsymbol{k}\right)	& = &	\left(A\boldsymbol{k}_{//}^{2}+ak_{z}^{2}+m_{0}\right)\sigma_{0}+Bk_{z}\left(k_{-}^{2}\sigma_{+}+k_{+}^{2}\sigma_{-}\right)\nonumber \\
& + &	\left[D\boldsymbol{k}_{//}^{2}+dk_{z}^{2}+m+Ck_{x}\left(k_{x}^{2}-3k_{y}^{2}\right)\right]\sigma_{z},
\end{eqnarray}
where $\sigma_{0}$ is the identity matrix and $\sigma_{\pm}=\sigma_{x}\pm i\sigma_{y}$ with $\sigma_{x,y,z}$ to be the Pauli matrices. The $\boldsymbol{k}_{//}$ is the in-plane momentum, the $A$, $B$, $C$, $D$, $a$, $d$, $m$, $m_0$ are fitting parameters from first-principles calculations.
By solving this Hamiltonian, we can obtain a pair of DWPs located at (0, 0, $k_z^c$ = $\pm\sqrt{-m/d}$) and a WNL in $k_z$ = 0 plane when the conditions $d{\cdot}m < 0$ and $D{\cdot}m < 0$ are satisfied.
The 3D energy dispersions of DWP at $k_z$ = $k_z^c$ and WNL in $k_z$ = 0 plane are plotted in Fig. \ref{FIG.3} (a) and (c). The DWP at $k_z$ = $k_z^c$ possesses a quadratic in-plane dispersion with chirality of +2. For any planes with fixed $k_z$ (except $k_z$ = 0, $\pm$ ${k_z^c}$), the system becomes an insulator and a $k_z$-dependent chern number can be well defined as shown in Fig. \ref{FIG.3} (b), where the +2 chern number appears in the region of  ${-k_z^c}<k_z<{k_z^c}$ that confirms it is a DWP. The Berry curvature distributions of DWPs are demonstrated in Figs. \ref{FIG.3} (e)-(f), in which the DWP work as a source or sink in the field of Berry curvature with triangular symmetry.
In Fig. \ref{FIG.3} (c), we notice the conduction and valence bands intersect around K point and produce a trigonal WNL structure at $k_z$ = 0 plane. This interesting trigonal warping effect of the WNL structure is induced by the $C_{3z}$ symmetry which is considered in $Ck_{x}\left(k_{x}^{2}-3k_{y}^{2}\right)$.
As shown in Fig. \ref{FIG.3} (b), the integral of Berry phase along an arbitrary closed path that interlocks the WNL indeed gives a $\pi$ Berry phase.

Taking a close look at the Hamiltonian, the first term breaks the electron-hole symmetry and brings the energy difference of band crossings, but this term has no concern with the topological nature. The second term provides couplings for WNL and DWPs when momentum $\boldsymbol{k}$ is neither at $k_z = 0$ plane nor at $k_z$ axis. The last term plays a vital role for the emergence of the nontrivial topological phase in Eu$_5$Bi$_3$. When $D{\cdot}m < 0$ and $d{\cdot}m < 0$, the band inversion happens, a DWPs and WNL fermion are produced concurrently to form a WNPL Fermion.

\begin{figure}[h]
	\includegraphics[width=8.8cm]{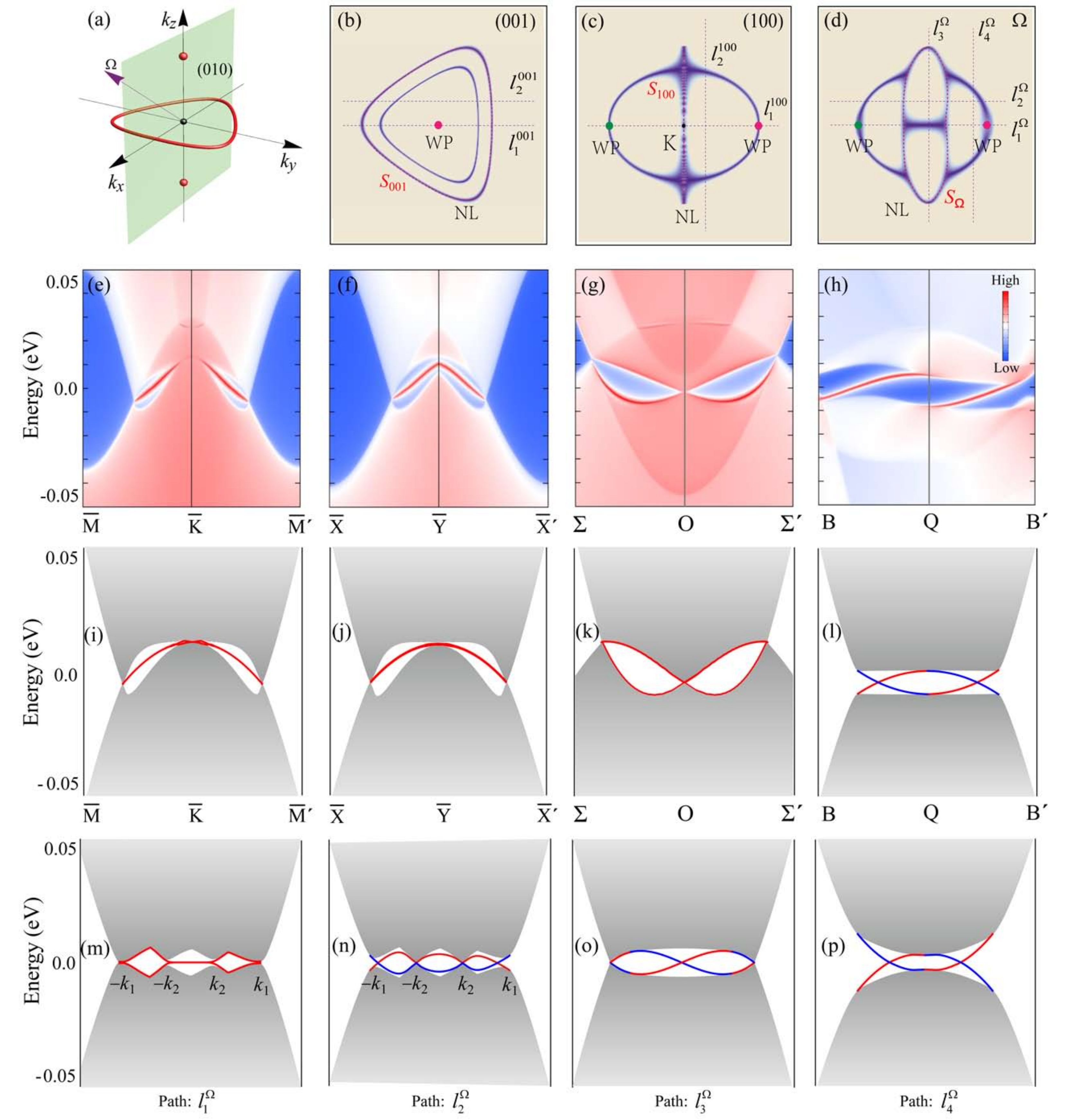}
	\caption{Surface states and Fermi arcs of WNPLSM. (a) Schematic illustrations of the WNPLSM in the 3D Brillouin zone. The normal direction of the $\Omega$ , i.e., the purple arrow is lying in the $k_{x}$-$k_{z}$ plane that is highlighted in lightgreen. (b), (c) and (d) show the calculated Fermi arcs of Eu$_5$Bi$_3$ for (001), (100) and $\Omega$ planes, respectively. Paths $l^{001}_1$, $l^{001}_2$, $l^{100}_1$, $l^{100}_2$ are equal to paths $\mathrm{\overline{M}}$-$\mathrm{\overline{K}}$-$\mathrm{\overline{M}'}$, $\mathrm{\overline{M}}$-$\mathrm{\overline{K}}$-$\mathrm{\overline{M}'}$, $\Sigma$-O-$\Sigma$$^{\prime}$, B-Q-B$^{\prime}$. (e)-(h) and (i)-(l) Surface band structures projected onto (001) and (100) planes calculated by wannier function and $\bm{k}\cdot\bm{p}$ Hamiltonian, respectively. (m)-(p) Surface band structures projected onto $\Omega$ plane calculated by $\bm{k}\cdot\bm{p}$ Hamiltonian.}\label{FIG.4}
\end{figure}

Based on the effective $\bm{k}\cdot\bm{p}$ model, we further explore the novel surface states and Fermi arcs of the WNPL fermion.
As shown in Figs. \ref{FIG.4} (b)-(d), we choose three representative projection planes (001), (100) and $\Omega$.
For the (001) surface as shown in Fig. \ref{FIG.4} (b), a WNL can be projected with maximal area (region $S_{001}$, red dashed line), and two DWPs are projected into a same point $\mathrm{\overline{K}}$ at the center of region $S_{001}$. The "drumhead" surface state from the WNL structure emerges inside region $S_{001}$ as shown in Figs. \ref{FIG.4} (i)-(j). Remarkably, we find that along $\mathrm{\overline{M}}$-$\mathrm{\overline{K}}$-$\mathrm{\overline{M}'}$ path, the "drumhead" surface state must cross $\mathrm{\overline{K}}$ point regardless of the surface circumstance. In other words, this "drumhead" surface state of the WNL is naturally pinned by the DWPs.

For (100) projection plane as shown in Fig. \ref{FIG.4} (c), the WNL is projected into a line segment (line $S_{100}$, red dashed line) and two DWPs are projected into both sides of line segment $S_{100}$. As shown in Fig. \ref{FIG.4} (l), along B-O-B$^{\prime}$ path, we can clearly see two branches of surface states with same chirality connect the conduction and valence bands, in good agreement with our calculated charity ($\pm$2) of DWPs. As shown in in Fig. \ref{FIG.4} (k), along $\Sigma$-O-$\Sigma$$^{\prime}$ path that connects two DWPs and crosses the WNL, we notice the surface states have to cross each other at WNL projection position, thus showing that the chiral surface states of DWPs are nontrivially pinned by the WNL structure. The Fermi surface of (100) surface state is displayed in Fig. \ref{FIG.4} (c). Two Fermi arcs stem from one DWP and cross both ends of the WNL projection region, and then gather together at the other DWP. This unique connection of surface Fermi arcs further demonstrates the coupling of DWPs and WNL in the WNPL fermion. These surface band structures of (001) and (100) projection planes are detailedly checked by our DFT calculations as shown in Figs. \ref{FIG.4} (e)-(h). Same morphologies of the surface states of WNPL fermion are revealed by DFT calculations, which matches well with that from $\bm{k}\cdot\bm{p}$ model.

For a general projection plane $\Omega$ as shown in Fig. \ref{FIG.4} (d), the WNL structure is projected into a small area (region $S_{\Omega}$, red dashed line) and two DWPs are projected into both sides of region $S_{\Omega}$. To get a global view of the coupled surface state, we chose four $k$-paths ($l_{1,2,3,4}^{\Omega}$) and corresponding surface states are listed in Figs. \ref{FIG.4} (m)-(p). First for $l_1^{\Omega}$ path in Fig. \ref{FIG.4} (m), two surface states from DWP emerge at $-k_1$ point and connect to the WNL at $-k_2$ point, similar to that in Fig. \ref{FIG.4} (k). Inside the WNL projection region $[-k_2, k_2]$, the surface state shows nearly flat dispersion. When  $l_1^{\Omega}$ path is parallelly moved to $l_2^{\Omega}$ as shown in Fig. \ref{FIG.4} (n), the WP's surface states still exist between $k_1$ and $k_2$, but the flat WNL's surface state becomes bended in region $[-k_2, k_2]$. This bending effect can be clearly observed along a vertical path $l_3^{\Omega}$ as shown in Fig. \ref{FIG.4} (o), where we can see the surface state on the top surface (red line) and on the bottom surface (blue line) possess opposite Fermi velocities. This means the surface state inside the WNL's projection region $S_{\Omega}$ also carries charity, denoted as chiral WNL surface states in WNPL structure. When $l_3^{\Omega}$ path is parallelly shifted to $l_4^{\Omega}$ outside $S_{\Omega}$, two chiral surface states from the DWPs connect the conduction and valence bands similar with that in Fig. \ref{FIG.4} (l). By plotting the Fermi surface of the surface state on $\Omega$ plane in Fig. \ref{FIG.4} (d), we reveal two Fermi arcs stemming from two DWPs connect to the boundary of $S_{\Omega}$. Inside $S_{\Omega}$ region, a new Fermi arc emerges which derived from the chiral surface state in WNPL structure.
Those unique chiral WNL surface states and fascinating links of surface Fermi arcs are significant characteristics that can discriminate the WNPL fermion from individual WP or WNL fermions.


The Weyl fermion in solid was initially proposed in a magnetic YIrO$_3$\cite{wan2011topological} but firstly experimentally observed in nonmagnetic TaAs families\cite{lv2015experimental}. In contrast with Weyl fermions in nonmagnetic materials, the magnetic Weyl fermions have the following advantages: (1) the least number of WPs, (2) the larger separation in reciprocal space for two WPs with opposite charities, (3) the giant anomalous Hall effect derived from the diverging Berry curvature of WPs\cite{kim2018large,yang2011quantum,wang2018large}.
The Weyl fermion proposed in ferromagnetic Eu$_5$Bi$_3$ perfectly meets these superiorities. Specifically, the system hosts a pair of WPs in momentum space with a big interval of 0.15 {\AA}$^{-1}$ and a large anomalous Hall conductivity of -260 ($\hbar$/$e$)($\Omega$cm)$^{-1}$ parallel to [001] direction is predicted.

In conclusion, we propose a new type TSM state, WNPL fermion, which is characterized by the intrinsic coexistence of WPs and WNL. We discover a SOC driven topological phase transition from magnetic TDP semimetal to WNPL semimetal in hexagonal Eu$_5$Bi$_3$.
The exotic surface state of WNPL fermion including pinned chiral surface state and unique linked Fermi arcs are demonstrated by both DFT calculations and effective $\bm{k}\cdot\bm{p}$ model.
Our work sheds light on discovering new topological semimetal states from diverse band-crossing dimensions, and point to an ideal material candidate Eu$_5$Bi$_3$ for future experimental realization. We believe this composite WNPL fermion can be further generalized to Weyl nodal point-surface and line-surface fermions.

\begin{acknowledgement}
This work is supported by the National Key R\&D Program of China (Grant No. 2016YFA0300600), the National Natural Science Foundation of China (Grants No. 11734003, No. 11574029), the Strategic Priority Research Program of Chinese Academy of Sciences (Grant No. XDB30000000), the Fundamental Research Funds for the Central Universities (Grant No.2017CX10007). B.T.F. also thanks Sichuan Normal University for financial support (No. 341829001). H.B.W. and D.S.M. also thank the supports from Graduate Technological Innovation Project of Beijing Institute of Technology (Grant No. 2018CX10028).

H.B.W. and D.S.M. contributed equally to this work.

\end{acknowledgement}

\textcolor{blue}{
\section{Supporting Information}
\begin{quote}
Supporting Information Available:
The computational methods and details for the constructions of the $\bm{k}\cdot\bm{p}$ effective Hamiltonian.
\end{quote}
}

\bibliography{Eubiref}

\providecommand*\mcitethebibliography{\thebibliography}
\csname @ifundefined\endcsname{endmcitethebibliography}
  {\let\endmcitethebibliography\endthebibliography}{}
\begin{mcitethebibliography}{66}
\providecommand*\natexlab[1]{#1}
\providecommand*\mciteSetBstSublistMode[1]{}
\providecommand*\mciteSetBstMaxWidthForm[2]{}
\providecommand*\mciteBstWouldAddEndPuncttrue
  {\def\EndOfBibitem{\unskip.}}
\providecommand*\mciteBstWouldAddEndPunctfalse
  {\let\EndOfBibitem\relax}
\providecommand*\mciteSetBstMidEndSepPunct[3]{}
\providecommand*\mciteSetBstSublistLabelBeginEnd[3]{}
\providecommand*\EndOfBibitem{}
\mciteSetBstSublistMode{f}
\mciteSetBstMaxWidthForm{subitem}{(\alph{mcitesubitemcount})}
\mciteSetBstSublistLabelBeginEnd
  {\mcitemaxwidthsubitemform\space}
  {\relax}
  {\relax}

\bibitem[Armitage et~al.(2018)Armitage, Mele, and Vishwanath]{armitage2018weyl}
Armitage,~N.; Mele,~E.; Vishwanath,~A. Weyl and Dirac semimetals in
  three-dimensional solids. \emph{Rev. Mod. Phys.} \textbf{2018}, \emph{90},
  015001\relax
\mciteBstWouldAddEndPuncttrue
\mciteSetBstMidEndSepPunct{\mcitedefaultmidpunct}
{\mcitedefaultendpunct}{\mcitedefaultseppunct}\relax
\EndOfBibitem
\bibitem[Young et~al.(2012)Young, Zaheer, Teo, Kane, Mele, and
  Rappe]{young2012dirac}
Young,~S.~M.; Zaheer,~S.; Teo,~J.~C.; Kane,~C.~L.; Mele,~E.~J.; Rappe,~A.~M.
  Dirac semimetal in three dimensions. \emph{Phys. Rev. Lett.} \textbf{2012},
  \emph{108}, 140405\relax
\mciteBstWouldAddEndPuncttrue
\mciteSetBstMidEndSepPunct{\mcitedefaultmidpunct}
{\mcitedefaultendpunct}{\mcitedefaultseppunct}\relax
\EndOfBibitem
\bibitem[Wang et~al.(2012)Wang, Sun, Chen, Franchini, Xu, Weng, Dai, and
  Fang]{wang2012dirac}
Wang,~Z.; Sun,~Y.; Chen,~X.-Q.; Franchini,~C.; Xu,~G.; Weng,~H.; Dai,~X.;
  Fang,~Z. Dirac semimetal and topological phase transitions in A$_{3}$Bi (A=
  Na, K, Rb). \emph{Phys. Rev. B} \textbf{2012}, \emph{85}, 195320\relax
\mciteBstWouldAddEndPuncttrue
\mciteSetBstMidEndSepPunct{\mcitedefaultmidpunct}
{\mcitedefaultendpunct}{\mcitedefaultseppunct}\relax
\EndOfBibitem
\bibitem[Xu et~al.(2015)Xu, Belopolski, Alidoust, Neupane, Bian, Zhang, Sankar,
  Chang, Yuan, Lee, et~al. others]{xu2015discovery}
Xu,~S.-Y.; Belopolski,~I.; Alidoust,~N.; Neupane,~M.; Bian,~G.; Zhang,~C.;
  Sankar,~R.; Chang,~G.; Yuan,~Z.; Lee,~C.-C. et~al.  Discovery of a Weyl
  fermion semimetal and topological Fermi arcs. \emph{Science} \textbf{2015},
  \emph{349}, 613--617\relax
\mciteBstWouldAddEndPuncttrue
\mciteSetBstMidEndSepPunct{\mcitedefaultmidpunct}
{\mcitedefaultendpunct}{\mcitedefaultseppunct}\relax
\EndOfBibitem
\bibitem[Chen et~al.(2015)Chen, Xie, Yang, Pan, Zhang, Cohen, and
  Zhang]{chen2015nl}
Chen,~Y.; Xie,~Y.; Yang,~S.~A.; Pan,~H.; Zhang,~F.; Cohen,~M.~L.; Zhang,~S.
  Nanostructured Carbon Allotropes with Weyl-like Loops and Points. \emph{Nano
  Lett.} \textbf{2015}, \emph{15}, 6974--6978\relax
\mciteBstWouldAddEndPuncttrue
\mciteSetBstMidEndSepPunct{\mcitedefaultmidpunct}
{\mcitedefaultendpunct}{\mcitedefaultseppunct}\relax
\EndOfBibitem
\bibitem[Weng et~al.(2016)Weng, Dai, and Fang]{weng2016topology}
Weng,~H.; Dai,~X.; Fang,~Z. Topological semimetals predicted from
  first-principles calculations. \emph{J. Phys.: Condens. Matter.}
  \textbf{2016}, \emph{28}, 303001\relax
\mciteBstWouldAddEndPuncttrue
\mciteSetBstMidEndSepPunct{\mcitedefaultmidpunct}
{\mcitedefaultendpunct}{\mcitedefaultseppunct}\relax
\EndOfBibitem
\bibitem[Chiu et~al.(2016)Chiu, Teo, Schnyder, and Ryu]{chiu2016classification}
Chiu,~C.-K.; Teo,~J.~C.; Schnyder,~A.~P.; Ryu,~S. Classification of topological
  quantum matter with symmetries. \emph{Rev. Mod. Phys.} \textbf{2016},
  \emph{88}, 035005\relax
\mciteBstWouldAddEndPuncttrue
\mciteSetBstMidEndSepPunct{\mcitedefaultmidpunct}
{\mcitedefaultendpunct}{\mcitedefaultseppunct}\relax
\EndOfBibitem
\bibitem[Huang et~al.(2018)Huang, Jin, Zhang, and Liu]{huang2018nl}
Huang,~H.; Jin,~K.-H.; Zhang,~S.; Liu,~F. Topological Electride Y$_{2}$C.
  \emph{Nano Lett.} \textbf{2018}, \emph{18}, 1972--1977\relax
\mciteBstWouldAddEndPuncttrue
\mciteSetBstMidEndSepPunct{\mcitedefaultmidpunct}
{\mcitedefaultendpunct}{\mcitedefaultseppunct}\relax
\EndOfBibitem
\bibitem[Hirayama et~al.(2018)Hirayama, Okugawa, and
  Murakami]{HirayamaMtop2018}
Hirayama,~M.; Okugawa,~R.; Murakami,~S. Topological semimetals studied by ab
  initio calculations. \emph{J. Phys. Soc. Jpn.} \textbf{2018}, \emph{87},
  041002\relax
\mciteBstWouldAddEndPuncttrue
\mciteSetBstMidEndSepPunct{\mcitedefaultmidpunct}
{\mcitedefaultendpunct}{\mcitedefaultseppunct}\relax
\EndOfBibitem
\bibitem[Jiang et~al.(2018)Jiang, Dun, Moon, Zhou, Koshino, Smirnov, and
  Jiang]{jiang2019nl}
Jiang,~Y.; Dun,~Z.; Moon,~S.; Zhou,~H.; Koshino,~M.; Smirnov,~D.; Jiang,~Z.
  Landau Quantization in Coupled Weyl Points: A Case Study of Semimetal NbP.
  \emph{Nano Lett.} \textbf{2018}, \emph{18}, 7726--7731\relax
\mciteBstWouldAddEndPuncttrue
\mciteSetBstMidEndSepPunct{\mcitedefaultmidpunct}
{\mcitedefaultendpunct}{\mcitedefaultseppunct}\relax
\EndOfBibitem
\bibitem[Zyuzin and Burkov(2012)Zyuzin, and Burkov]{zyuzin2012topological}
Zyuzin,~A.; Burkov,~A. Topological response in Weyl semimetals and the chiral
  anomaly. \emph{Phys. Rev. B} \textbf{2012}, \emph{86}, 115133\relax
\mciteBstWouldAddEndPuncttrue
\mciteSetBstMidEndSepPunct{\mcitedefaultmidpunct}
{\mcitedefaultendpunct}{\mcitedefaultseppunct}\relax
\EndOfBibitem
\bibitem[Son and Spivak(2013)Son, and Spivak]{son2013chiral}
Son,~D.; Spivak,~B. Chiral anomaly and classical negative magnetoresistance of
  Weyl metals. \emph{Phys. Rev. B} \textbf{2013}, \emph{88}, 104412\relax
\mciteBstWouldAddEndPuncttrue
\mciteSetBstMidEndSepPunct{\mcitedefaultmidpunct}
{\mcitedefaultendpunct}{\mcitedefaultseppunct}\relax
\EndOfBibitem
\bibitem[Gorbar et~al.(2014)Gorbar, Miransky, and Shovkovy]{gorbar2014chiral}
Gorbar,~E.; Miransky,~V.; Shovkovy,~I. Chiral anomaly, dimensional reduction,
  and magnetoresistivity of Weyl and Dirac semimetals. \emph{Phys. Rev. B}
  \textbf{2014}, \emph{89}, 085126\relax
\mciteBstWouldAddEndPuncttrue
\mciteSetBstMidEndSepPunct{\mcitedefaultmidpunct}
{\mcitedefaultendpunct}{\mcitedefaultseppunct}\relax
\EndOfBibitem
\bibitem[Huang et~al.(2015)Huang, Zhao, Long, Wang, Chen, Yang, Liang, Xue,
  Weng, Fang, et~al. others]{huang2015observation}
Huang,~X.; Zhao,~L.; Long,~Y.; Wang,~P.; Chen,~D.; Yang,~Z.; Liang,~H.;
  Xue,~M.; Weng,~H.; Fang,~Z. et~al.  Observation of the chiral-anomaly-induced
  negative magnetoresistance in 3D Weyl semimetal TaAs. \emph{Phys. Rev. X}
  \textbf{2015}, \emph{5}, 031023\relax
\mciteBstWouldAddEndPuncttrue
\mciteSetBstMidEndSepPunct{\mcitedefaultmidpunct}
{\mcitedefaultendpunct}{\mcitedefaultseppunct}\relax
\EndOfBibitem
\bibitem[Heikkil{\"a} et~al.(2011)Heikkil{\"a}, Kopnin, and
  Volovik]{heikkila2011flat}
Heikkil{\"a},~T.~T.; Kopnin,~N.~B.; Volovik,~G.~E. Flat bands in topological
  media. \emph{JETP Lett.} \textbf{2011}, \emph{94}, 233\relax
\mciteBstWouldAddEndPuncttrue
\mciteSetBstMidEndSepPunct{\mcitedefaultmidpunct}
{\mcitedefaultendpunct}{\mcitedefaultseppunct}\relax
\EndOfBibitem
\bibitem[Qi et~al.(2016)Qi, Naumov, Ali, Rajamathi, Schnelle, Barkalov,
  Hanfland, Wu, Shekhar, Sun, et~al. others]{qi2016superconductivity}
Qi,~Y.; Naumov,~P.~G.; Ali,~M.~N.; Rajamathi,~C.~R.; Schnelle,~W.;
  Barkalov,~O.; Hanfland,~M.; Wu,~S.-C.; Shekhar,~C.; Sun,~Y. et~al.
  Superconductivity in Weyl semimetal candidate MoTe$_{2}$. \emph{Nat. Commun.}
  \textbf{2016}, \emph{7}, 11038\relax
\mciteBstWouldAddEndPuncttrue
\mciteSetBstMidEndSepPunct{\mcitedefaultmidpunct}
{\mcitedefaultendpunct}{\mcitedefaultseppunct}\relax
\EndOfBibitem
\bibitem[Chan et~al.(2016)Chan, Chiu, Chou, and Schnyder]{chan20163}
Chan,~Y.-H.; Chiu,~C.-K.; Chou,~M.; Schnyder,~A.~P. Ca$_{3}$P$_{2}$ and other
  topological semimetals with line nodes and drumhead surface states.
  \emph{Phys. Rev. B} \textbf{2016}, \emph{93}, 205132\relax
\mciteBstWouldAddEndPuncttrue
\mciteSetBstMidEndSepPunct{\mcitedefaultmidpunct}
{\mcitedefaultendpunct}{\mcitedefaultseppunct}\relax
\EndOfBibitem
\bibitem[Wang and Nandkishore(2017)Wang, and Nandkishore]{wang2017topological}
Wang,~Y.; Nandkishore,~R.~M. Topological surface superconductivity in doped
  Weyl loop materials. \emph{Phys. Rev. B} \textbf{2017}, \emph{95},
  060506\relax
\mciteBstWouldAddEndPuncttrue
\mciteSetBstMidEndSepPunct{\mcitedefaultmidpunct}
{\mcitedefaultendpunct}{\mcitedefaultseppunct}\relax
\EndOfBibitem
\bibitem[Br{\"u}ne et~al.(2011)Br{\"u}ne, Liu, Novik, Hankiewicz, Buhmann,
  Chen, Qi, Shen, Zhang, and Molenkamp]{brune2011quantum}
Br{\"u}ne,~C.; Liu,~C.; Novik,~E.; Hankiewicz,~E.; Buhmann,~H.; Chen,~Y.;
  Qi,~X.; Shen,~Z.; Zhang,~S.; Molenkamp,~L. Quantum Hall effect from the
  topological surface states of strained bulk HgTe. \emph{Phys. Rev. Lett.}
  \textbf{2011}, \emph{106}, 126803\relax
\mciteBstWouldAddEndPuncttrue
\mciteSetBstMidEndSepPunct{\mcitedefaultmidpunct}
{\mcitedefaultendpunct}{\mcitedefaultseppunct}\relax
\EndOfBibitem
\bibitem[Wang et~al.(2017)Wang, Sun, Lu, and Xie]{wang20173d}
Wang,~C.; Sun,~H.-P.; Lu,~H.-Z.; Xie,~X. 3D quantum Hall effect of Fermi arcs
  in topological semimetals. \emph{Phys. Rev. Lett.} \textbf{2017}, \emph{119},
  136806\relax
\mciteBstWouldAddEndPuncttrue
\mciteSetBstMidEndSepPunct{\mcitedefaultmidpunct}
{\mcitedefaultendpunct}{\mcitedefaultseppunct}\relax
\EndOfBibitem
\bibitem[Zhang et~al.(2018)Zhang, Zhang, Yuan, Lu, Zhang, Narayan, Liu, Zhang,
  Ni, Liu, Choi, Suslov, Sanvito, Pi, Lu, Potter, and Xiu]{zhang2018quantum}
Zhang,~C.; Zhang,~Y.; Yuan,~X.; Lu,~S.; Zhang,~J.; Narayan,~A.; Liu,~Y.;
  Zhang,~H.; Ni,~Z.; Liu,~R. et~al.  Quantum Hall effect based on Weyl orbits
  in Cd$_{3}$As$_{2}$. \emph{Nature} \textbf{2018}, 1\relax
\mciteBstWouldAddEndPuncttrue
\mciteSetBstMidEndSepPunct{\mcitedefaultmidpunct}
{\mcitedefaultendpunct}{\mcitedefaultseppunct}\relax
\EndOfBibitem
\bibitem[Burkov and Balents(2011)Burkov, and Balents]{burkov2011weyl}
Burkov,~A.; Balents,~L. Weyl semimetal in a topological insulator multilayer.
  \emph{Phys. Rev. Lett.} \textbf{2011}, \emph{107}, 127205\relax
\mciteBstWouldAddEndPuncttrue
\mciteSetBstMidEndSepPunct{\mcitedefaultmidpunct}
{\mcitedefaultendpunct}{\mcitedefaultseppunct}\relax
\EndOfBibitem
\bibitem[Lv et~al.(2015)Lv, Weng, Fu, Wang, Miao, Ma, Richard, Huang, Zhao,
  Chen, et~al. others]{lv2015experimental}
Lv,~B.; Weng,~H.; Fu,~B.; Wang,~X.; Miao,~H.; Ma,~J.; Richard,~P.; Huang,~X.;
  Zhao,~L.; Chen,~G. et~al.  Experimental discovery of Weyl semimetal TaAs.
  \emph{Phys. Rev. X} \textbf{2015}, \emph{5}, 031013\relax
\mciteBstWouldAddEndPuncttrue
\mciteSetBstMidEndSepPunct{\mcitedefaultmidpunct}
{\mcitedefaultendpunct}{\mcitedefaultseppunct}\relax
\EndOfBibitem
\bibitem[Soluyanov et~al.(2015)Soluyanov, Gresch, Wang, Wu, Troyer, Dai, and
  Bernevig]{soluyanov2015type}
Soluyanov,~A.~A.; Gresch,~D.; Wang,~Z.; Wu,~Q.; Troyer,~M.; Dai,~X.;
  Bernevig,~B.~A. Type-II weyl semimetals. \emph{Nature} \textbf{2015},
  \emph{527}, 495\relax
\mciteBstWouldAddEndPuncttrue
\mciteSetBstMidEndSepPunct{\mcitedefaultmidpunct}
{\mcitedefaultendpunct}{\mcitedefaultseppunct}\relax
\EndOfBibitem
\bibitem[Burkov et~al.(2011)Burkov, Hook, and Balents]{burkov2011topological}
Burkov,~A.; Hook,~M.; Balents,~L. Topological nodal semimetals. \emph{Phys.
  Rev. B} \textbf{2011}, \emph{84}, 235126\relax
\mciteBstWouldAddEndPuncttrue
\mciteSetBstMidEndSepPunct{\mcitedefaultmidpunct}
{\mcitedefaultendpunct}{\mcitedefaultseppunct}\relax
\EndOfBibitem
\bibitem[Mullen et~al.(2015)Mullen, Uchoa, and Glatzhofer]{mullen2015line}
Mullen,~K.; Uchoa,~B.; Glatzhofer,~D.~T. Line of Dirac nodes in hyperhoneycomb
  lattices. \emph{Phys. Rev. Lett.} \textbf{2015}, \emph{115}, 026403\relax
\mciteBstWouldAddEndPuncttrue
\mciteSetBstMidEndSepPunct{\mcitedefaultmidpunct}
{\mcitedefaultendpunct}{\mcitedefaultseppunct}\relax
\EndOfBibitem
\bibitem[Fang et~al.(2016)Fang, Weng, Dai, and Fang]{fang2016topological}
Fang,~C.; Weng,~H.; Dai,~X.; Fang,~Z. Topological nodal line semimetals.
  \emph{Chin. Phys. B} \textbf{2016}, \emph{25}, 117106\relax
\mciteBstWouldAddEndPuncttrue
\mciteSetBstMidEndSepPunct{\mcitedefaultmidpunct}
{\mcitedefaultendpunct}{\mcitedefaultseppunct}\relax
\EndOfBibitem
\bibitem[Hu et~al.(2016)Hu, Tang, Liu, Liu, Zhu, Graf, Myhro, Tran, Lau, Wei,
  et~al. others]{hu2016evidence}
Hu,~J.; Tang,~Z.; Liu,~J.; Liu,~X.; Zhu,~Y.; Graf,~D.; Myhro,~K.; Tran,~S.;
  Lau,~C.~N.; Wei,~J. et~al.  Evidence of topological nodal-line fermions in
  ZrSiSe and ZrSiTe. \emph{Phys. Rev. Lett.} \textbf{2016}, \emph{117},
  016602\relax
\mciteBstWouldAddEndPuncttrue
\mciteSetBstMidEndSepPunct{\mcitedefaultmidpunct}
{\mcitedefaultendpunct}{\mcitedefaultseppunct}\relax
\EndOfBibitem
\bibitem[Li et~al.(2017)Li, Yu, Liu, Guan, Wang, Zhang, Yao, and
  Yang]{li2017type}
Li,~S.; Yu,~Z.-M.; Liu,~Y.; Guan,~S.; Wang,~S.-S.; Zhang,~X.; Yao,~Y.;
  Yang,~S.~A. Type-II nodal loops: Theory and material realization. \emph{Phys.
  Rev. B} \textbf{2017}, \emph{96}, 081106\relax
\mciteBstWouldAddEndPuncttrue
\mciteSetBstMidEndSepPunct{\mcitedefaultmidpunct}
{\mcitedefaultendpunct}{\mcitedefaultseppunct}\relax
\EndOfBibitem
\bibitem[Fang et~al.(2015)Fang, Chen, Kee, and Fu]{fang2015topological}
Fang,~C.; Chen,~Y.; Kee,~H.-Y.; Fu,~L. Topological nodal line semimetals with
  and without spin-orbital coupling. \emph{Phys. Rev. B} \textbf{2015},
  \emph{92}, 081201\relax
\mciteBstWouldAddEndPuncttrue
\mciteSetBstMidEndSepPunct{\mcitedefaultmidpunct}
{\mcitedefaultendpunct}{\mcitedefaultseppunct}\relax
\EndOfBibitem
\bibitem[Zhong et~al.(2016)Zhong, Chen, Xie, Yang, Cohen, and
  Zhang]{zhong2016towards}
Zhong,~C.; Chen,~Y.; Xie,~Y.; Yang,~S.~A.; Cohen,~M.~L.; Zhang,~S. Towards
  three-dimensional Weyl-surface semimetals in graphene networks.
  \emph{Nanoscale} \textbf{2016}, \emph{8}, 7232--7239\relax
\mciteBstWouldAddEndPuncttrue
\mciteSetBstMidEndSepPunct{\mcitedefaultmidpunct}
{\mcitedefaultendpunct}{\mcitedefaultseppunct}\relax
\EndOfBibitem
\bibitem[Liang et~al.(2016)Liang, Zhou, Yu, Wang, and Weng]{liang2016node}
Liang,~Q.-F.; Zhou,~J.; Yu,~R.; Wang,~Z.; Weng,~H. Node-surface and node-line
  fermions from nonsymmorphic lattice symmetries. \emph{Phys. Rev. B}
  \textbf{2016}, \emph{93}, 085427\relax
\mciteBstWouldAddEndPuncttrue
\mciteSetBstMidEndSepPunct{\mcitedefaultmidpunct}
{\mcitedefaultendpunct}{\mcitedefaultseppunct}\relax
\EndOfBibitem
\bibitem[Bzdu{\v{s}}ek and Sigrist(2017)Bzdu{\v{s}}ek, and
  Sigrist]{bzduvsek2017robust}
Bzdu{\v{s}}ek,~T.; Sigrist,~M. Robust doubly charged nodal lines and nodal
  surfaces in centrosymmetric systems. \emph{Phys. Rev. B} \textbf{2017},
  \emph{96}, 155105\relax
\mciteBstWouldAddEndPuncttrue
\mciteSetBstMidEndSepPunct{\mcitedefaultmidpunct}
{\mcitedefaultendpunct}{\mcitedefaultseppunct}\relax
\EndOfBibitem
\bibitem[Weng et~al.(2015)Weng, Fang, Fang, Bernevig, and Dai]{weng2015weyl}
Weng,~H.; Fang,~C.; Fang,~Z.; Bernevig,~B.~A.; Dai,~X. Weyl semimetal phase in
  noncentrosymmetric transition-metal monophosphides. \emph{Phys. Rev. X}
  \textbf{2015}, \emph{5}, 011029\relax
\mciteBstWouldAddEndPuncttrue
\mciteSetBstMidEndSepPunct{\mcitedefaultmidpunct}
{\mcitedefaultendpunct}{\mcitedefaultseppunct}\relax
\EndOfBibitem
\bibitem[Yan and Felser(2017)Yan, and Felser]{yan2017topological}
Yan,~B.; Felser,~C. Topological materials: Weyl semimetals. \emph{Annu. Rev.
  Condens. Matter Phys.} \textbf{2017}, \emph{8}, 337--354\relax
\mciteBstWouldAddEndPuncttrue
\mciteSetBstMidEndSepPunct{\mcitedefaultmidpunct}
{\mcitedefaultendpunct}{\mcitedefaultseppunct}\relax
\EndOfBibitem
\bibitem[Burkov(2018)]{burkov2018weyl}
Burkov,~A. Weyl metals. \emph{Annu. Rev. Condens. Matter Phys.} \textbf{2018},
  \emph{9}, 359--378\relax
\mciteBstWouldAddEndPuncttrue
\mciteSetBstMidEndSepPunct{\mcitedefaultmidpunct}
{\mcitedefaultendpunct}{\mcitedefaultseppunct}\relax
\EndOfBibitem
\bibitem[Wang et~al.(2018)Wang, Zhao, Jin, Du, Zhao, Xu, and
  Tong]{wang2018nodal}
Wang,~R.; Zhao,~J.; Jin,~Y.; Du,~Y.; Zhao,~Y.; Xu,~H.; Tong,~S. Nodal line
  fermions in magnetic oxides. \emph{Phys. Rev. B} \textbf{2018}, \emph{97},
  241111\relax
\mciteBstWouldAddEndPuncttrue
\mciteSetBstMidEndSepPunct{\mcitedefaultmidpunct}
{\mcitedefaultendpunct}{\mcitedefaultseppunct}\relax
\EndOfBibitem
\bibitem[Feng et~al.(2019)Feng, Zhang, Feng, Fu, Wu, Miyamoto, He, Chen, Wu,
  Shimada, Okuda, and Yao]{fengbaojie2019}
Feng,~B.; Zhang,~R.-W.; Feng,~Y.; Fu,~B.; Wu,~S.; Miyamoto,~K.; He,~S.;
  Chen,~L.; Wu,~K.; Shimada,~K. et~al.  Discovery of Weyl nodal lines in a
  single-layer ferromagnet. \emph{{arXiv}:1901.01429} \textbf{2019}, \relax
\mciteBstWouldAddEndPunctfalse
\mciteSetBstMidEndSepPunct{\mcitedefaultmidpunct}
{}{\mcitedefaultseppunct}\relax
\EndOfBibitem
\bibitem[Wan et~al.(2011)Wan, Turner, Vishwanath, and
  Savrasov]{wan2011topological}
Wan,~X.; Turner,~A.~M.; Vishwanath,~A.; Savrasov,~S.~Y. Topological semimetal
  and Fermi-arc surface states in the electronic structure of pyrochlore
  iridates. \emph{Phys. Rev. B} \textbf{2011}, \emph{83}, 205101\relax
\mciteBstWouldAddEndPuncttrue
\mciteSetBstMidEndSepPunct{\mcitedefaultmidpunct}
{\mcitedefaultendpunct}{\mcitedefaultseppunct}\relax
\EndOfBibitem
\bibitem[Hasan et~al.(2017)Hasan, Xu, Belopolski, and
  Huang]{hasan2017discovery}
Hasan,~M.~Z.; Xu,~S.-Y.; Belopolski,~I.; Huang,~S.-M. Discovery of Weyl fermion
  semimetals and topological Fermi arc states. \emph{Annu. Rev. Condens. Matter
  Phys.} \textbf{2017}, \emph{8}, 289--309\relax
\mciteBstWouldAddEndPuncttrue
\mciteSetBstMidEndSepPunct{\mcitedefaultmidpunct}
{\mcitedefaultendpunct}{\mcitedefaultseppunct}\relax
\EndOfBibitem
\bibitem[Belopolski et~al.(2016)Belopolski, Xu, Ishida, Pan, Yu, Sanchez,
  Zheng, Neupane, Alidoust, Chang, et~al. others]{belopolski2016fermi}
Belopolski,~I.; Xu,~S.-Y.; Ishida,~Y.; Pan,~X.; Yu,~P.; Sanchez,~D.~S.;
  Zheng,~H.; Neupane,~M.; Alidoust,~N.; Chang,~G. et~al.  Fermi arc electronic
  structure and Chern numbers in the type-II Weyl semimetal candidate
  Mo$_{x}$W$_{1-x}$Te$_{2}$. \emph{Phys. Rev. B} \textbf{2016}, \emph{94},
  085127\relax
\mciteBstWouldAddEndPuncttrue
\mciteSetBstMidEndSepPunct{\mcitedefaultmidpunct}
{\mcitedefaultendpunct}{\mcitedefaultseppunct}\relax
\EndOfBibitem
\bibitem[Deng et~al.(2016)Deng, Wan, Deng, Zhang, Ding, Wang, Yan, Huang,
  Zhang, Xu, et~al. others]{deng2016experimental}
Deng,~K.; Wan,~G.; Deng,~P.; Zhang,~K.; Ding,~S.; Wang,~E.; Yan,~M.; Huang,~H.;
  Zhang,~H.; Xu,~Z. et~al.  Experimental observation of topological Fermi arcs
  in type-II Weyl semimetal MoTe$_{2}$. \emph{Nat. Phys.} \textbf{2016},
  \emph{12}, 1105\relax
\mciteBstWouldAddEndPuncttrue
\mciteSetBstMidEndSepPunct{\mcitedefaultmidpunct}
{\mcitedefaultendpunct}{\mcitedefaultseppunct}\relax
\EndOfBibitem
\bibitem[Fang et~al.(2012)Fang, Gilbert, Dai, and Bernevig]{fang2012multi}
Fang,~C.; Gilbert,~M.~J.; Dai,~X.; Bernevig,~B.~A. Multi-Weyl topological
  semimetals stabilized by point group symmetry. \emph{Phys. Rev. Lett.}
  \textbf{2012}, \emph{108}, 266802\relax
\mciteBstWouldAddEndPuncttrue
\mciteSetBstMidEndSepPunct{\mcitedefaultmidpunct}
{\mcitedefaultendpunct}{\mcitedefaultseppunct}\relax
\EndOfBibitem
\bibitem[Jian and Yao(2015)Jian, and Yao]{jian2015correlated}
Jian,~S.-K.; Yao,~H. Correlated double-Weyl semimetals with Coulomb
  interactions: Possible applications to HgCr$_{2}$Se$_{4}$ and SrSi$_{2}$.
  \emph{Phys. Rev. B} \textbf{2015}, \emph{92}, 045121\relax
\mciteBstWouldAddEndPuncttrue
\mciteSetBstMidEndSepPunct{\mcitedefaultmidpunct}
{\mcitedefaultendpunct}{\mcitedefaultseppunct}\relax
\EndOfBibitem
\bibitem[Huang et~al.(2016)Huang, Xu, Belopolski, Lee, Chang, Chang, Wang,
  Alidoust, Bian, Neupane, et~al. others]{huang2016new}
Huang,~S.-M.; Xu,~S.-Y.; Belopolski,~I.; Lee,~C.-C.; Chang,~G.; Chang,~T.-R.;
  Wang,~B.; Alidoust,~N.; Bian,~G.; Neupane,~M. et~al.  New type of Weyl
  semimetal with quadratic double Weyl fermions. \emph{Proc. Natl. Acad. Sci.
  USA} \textbf{2016}, \emph{113}, 1180--1185\relax
\mciteBstWouldAddEndPuncttrue
\mciteSetBstMidEndSepPunct{\mcitedefaultmidpunct}
{\mcitedefaultendpunct}{\mcitedefaultseppunct}\relax
\EndOfBibitem
\bibitem[Tsirkin et~al.()Tsirkin, Souza, and Vanderbilt]{Tsirkin2017Composite}
Tsirkin,~S.~S.; Souza,~I.; Vanderbilt,~D. Composite Weyl nodes stabilized by
  screw symmetry with and without time reversal. \emph{Phys. Rev. B} \emph{96},
  045102\relax
\mciteBstWouldAddEndPuncttrue
\mciteSetBstMidEndSepPunct{\mcitedefaultmidpunct}
{\mcitedefaultendpunct}{\mcitedefaultseppunct}\relax
\EndOfBibitem
\bibitem[Zaheer et~al.(2013)Zaheer, Young, Cellucci, Teo, Kane, Mele, and
  Rappe]{zaheer2013spin}
Zaheer,~S.; Young,~S.~M.; Cellucci,~D.; Teo,~J.~C.; Kane,~C.~L.; Mele,~E.~J.;
  Rappe,~A.~M. Spin texture on the Fermi surface of tensile-strained HgTe.
  \emph{Phys. Rev. B} \textbf{2013}, \emph{87}, 045202\relax
\mciteBstWouldAddEndPuncttrue
\mciteSetBstMidEndSepPunct{\mcitedefaultmidpunct}
{\mcitedefaultendpunct}{\mcitedefaultseppunct}\relax
\EndOfBibitem
\bibitem[Weng et~al.(2016)Weng, Fang, Fang, and Dai]{weng2016topological}
Weng,~H.; Fang,~C.; Fang,~Z.; Dai,~X. Topological semimetals with triply
  degenerate nodal points in $\theta$-phase tantalum nitride. \emph{Phys. Rev.
  B} \textbf{2016}, \emph{93}, 241202\relax
\mciteBstWouldAddEndPuncttrue
\mciteSetBstMidEndSepPunct{\mcitedefaultmidpunct}
{\mcitedefaultendpunct}{\mcitedefaultseppunct}\relax
\EndOfBibitem
\bibitem[Winkler et~al.(2016)Winkler, Wu, Troyer, Krogstrup, and
  Soluyanov]{winkler2016topological}
Winkler,~G.~W.; Wu,~Q.; Troyer,~M.; Krogstrup,~P.; Soluyanov,~A.~A. Topological
  phases in InAs$_{1-x}$Sb$_{x}$: from novel topological semimetal to Majorana
  wire. \emph{Phys. Rev. Lett.} \textbf{2016}, \emph{117}, 076403\relax
\mciteBstWouldAddEndPuncttrue
\mciteSetBstMidEndSepPunct{\mcitedefaultmidpunct}
{\mcitedefaultendpunct}{\mcitedefaultseppunct}\relax
\EndOfBibitem
\bibitem[Zhu et~al.(2016)Zhu, Winkler, Wu, Li, and Soluyanov]{zhu2016triple}
Zhu,~Z.; Winkler,~G.~W.; Wu,~Q.; Li,~J.; Soluyanov,~A.~A. Triple point
  topological metals. \emph{Phys. Rev. X} \textbf{2016}, \emph{6}, 031003\relax
\mciteBstWouldAddEndPuncttrue
\mciteSetBstMidEndSepPunct{\mcitedefaultmidpunct}
{\mcitedefaultendpunct}{\mcitedefaultseppunct}\relax
\EndOfBibitem
\bibitem[Yang et~al.(2017)Yang, Yu, Parkin, Felser, Liu, and
  Yan]{yang2017prediction}
Yang,~H.; Yu,~J.; Parkin,~S.~S.; Felser,~C.; Liu,~C.-X.; Yan,~B. Prediction of
  triple point fermions in simple half-Heusler topological insulators.
  \emph{Phys. Rev. Lett.} \textbf{2017}, \emph{119}, 136401\relax
\mciteBstWouldAddEndPuncttrue
\mciteSetBstMidEndSepPunct{\mcitedefaultmidpunct}
{\mcitedefaultendpunct}{\mcitedefaultseppunct}\relax
\EndOfBibitem
\bibitem[Gao et~al.(2018)Gao, Zhu, Zheng, Wu, Zhang, Xi, Zhang, Zhang, Hao,
  Ning, et~al. others]{gao2018possible}
Gao,~W.; Zhu,~X.; Zheng,~F.; Wu,~M.; Zhang,~J.; Xi,~C.; Zhang,~P.; Zhang,~Y.;
  Hao,~N.; Ning,~W. et~al.  A possible candidate for triply degenerate point
  fermions in trigonal layered PtBi$_{2}$. \emph{Nat. Commun.} \textbf{2018},
  \emph{9}\relax
\mciteBstWouldAddEndPuncttrue
\mciteSetBstMidEndSepPunct{\mcitedefaultmidpunct}
{\mcitedefaultendpunct}{\mcitedefaultseppunct}\relax
\EndOfBibitem
\bibitem[Zhang et~al.(2017)Zhang, Yu, Guo, Shi, Zhang, and Yao]{zhang2017jpcl}
Zhang,~T.-T.; Yu,~Z.-M.; Guo,~W.; Shi,~D.; Zhang,~G.; Yao,~Y. From Type-II
  Triply Degenerate Nodal Points and Three-Band Nodal Rings to Type-II Dirac
  Points in Centrosymmetric Zirconium Oxide. \emph{J. Phys. Chem. Lett.}
  \textbf{2017}, \emph{8}, 5792--5797\relax
\mciteBstWouldAddEndPuncttrue
\mciteSetBstMidEndSepPunct{\mcitedefaultmidpunct}
{\mcitedefaultendpunct}{\mcitedefaultseppunct}\relax
\EndOfBibitem
\bibitem[Neupane et~al.(2016)Neupane, Belopolski, Hosen, Sanchez, Sankar,
  Szlawska, Xu, Dimitri, Dhakal, Maldonado, et~al.
  others]{neupane2016observation}
Neupane,~M.; Belopolski,~I.; Hosen,~M.~M.; Sanchez,~D.~S.; Sankar,~R.;
  Szlawska,~M.; Xu,~S.-Y.; Dimitri,~K.; Dhakal,~N.; Maldonado,~P. et~al.
  Observation of topological nodal fermion semimetal phase in ZrSiS.
  \emph{Phys. Rev. B} \textbf{2016}, \emph{93}, 201104\relax
\mciteBstWouldAddEndPuncttrue
\mciteSetBstMidEndSepPunct{\mcitedefaultmidpunct}
{\mcitedefaultendpunct}{\mcitedefaultseppunct}\relax
\EndOfBibitem
\bibitem[Bian et~al.(2016)Bian, Chang, Sankar, Xu, Zheng, Neupert, Chiu, Huang,
  Chang, Belopolski, et~al. others]{bian2016topological}
Bian,~G.; Chang,~T.-R.; Sankar,~R.; Xu,~S.-Y.; Zheng,~H.; Neupert,~T.;
  Chiu,~C.-K.; Huang,~S.-M.; Chang,~G.; Belopolski,~I. et~al.  Topological
  nodal-line fermions in spin-orbit metal PbTaSe$_{2}$. \emph{Nat. Commun.}
  \textbf{2016}, \emph{7}, 10556\relax
\mciteBstWouldAddEndPuncttrue
\mciteSetBstMidEndSepPunct{\mcitedefaultmidpunct}
{\mcitedefaultendpunct}{\mcitedefaultseppunct}\relax
\EndOfBibitem
\bibitem[Wang et~al.(2016)Wang, Pan, Gao, Yu, Jiang, Zhang, Zuo, Zhang, Wei,
  Niu, et~al. others]{wang2016evidence}
Wang,~X.; Pan,~X.; Gao,~M.; Yu,~J.; Jiang,~J.; Zhang,~J.; Zuo,~H.; Zhang,~M.;
  Wei,~Z.; Niu,~W. et~al.  Evidence of both surface and bulk Dirac bands and
  anisotropic nonsaturating magnetoresistance in ZrSiS. \emph{Adv. Electron.
  Mater.} \textbf{2016}, \emph{2}, 1600228\relax
\mciteBstWouldAddEndPuncttrue
\mciteSetBstMidEndSepPunct{\mcitedefaultmidpunct}
{\mcitedefaultendpunct}{\mcitedefaultseppunct}\relax
\EndOfBibitem
\bibitem[Gao et~al.(2016)Gao, Hua, Zhang, and Zhang]{gao2016classification}
Gao,~Z.; Hua,~M.; Zhang,~H.; Zhang,~X. Classification of stable Dirac and Weyl
  semimetals with reflection and rotational symmetry. \emph{Phys. Rev. B}
  \textbf{2016}, \emph{93}, 205109\relax
\mciteBstWouldAddEndPuncttrue
\mciteSetBstMidEndSepPunct{\mcitedefaultmidpunct}
{\mcitedefaultendpunct}{\mcitedefaultseppunct}\relax
\EndOfBibitem
\bibitem[Yang et~al.(2018)Yang, Yang, Derunova, Parkin, Yan, and
  Ali]{yang2017symmetry}
Yang,~S.-Y.; Yang,~H.; Derunova,~E.; Parkin,~S. S.~P.; Yan,~B.; Ali,~M.~N.
  Symmetry Demanded Topological Nodal-line Materials. \emph{Adv. Phys.: X}
  \textbf{2018}, \emph{3}, 1414631\relax
\mciteBstWouldAddEndPuncttrue
\mciteSetBstMidEndSepPunct{\mcitedefaultmidpunct}
{\mcitedefaultendpunct}{\mcitedefaultseppunct}\relax
\EndOfBibitem
\bibitem[Kim et~al.(2018)Kim, Seo, Lee, Ko, Kim, Jang, Ok, Lee, Jo, Kang,
  et~al. others]{kim2018large}
Kim,~K.; Seo,~J.; Lee,~E.; Ko,~K.-T.; Kim,~B.; Jang,~B.~G.; Ok,~J.~M.; Lee,~J.;
  Jo,~Y.~J.; Kang,~W. et~al.  Large anomalous Hall current induced by
  topological nodal lines in a ferromagnetic van der Waals semimetal.
  \emph{Nat. Mater.} \textbf{2018}, \emph{17}, 794\relax
\mciteBstWouldAddEndPuncttrue
\mciteSetBstMidEndSepPunct{\mcitedefaultmidpunct}
{\mcitedefaultendpunct}{\mcitedefaultseppunct}\relax
\EndOfBibitem
\bibitem[Wang et~al.(2018)Wang, Xu, Lou, Liu, Li, Huang, Shen, Weng, Wang, and
  Lei]{wang2018large}
Wang,~Q.; Xu,~Y.; Lou,~R.; Liu,~Z.; Li,~M.; Huang,~Y.; Shen,~D.; Weng,~H.;
  Wang,~S.; Lei,~H. Large intrinsic anomalous Hall effect in half-metallic
  ferromagnet Co$_{3}$Sn$_{2}$S$_{2}$ with magnetic Weyl fermions. \emph{Nat.
  Commun.} \textbf{2018}, \emph{9}, 3681\relax
\mciteBstWouldAddEndPuncttrue
\mciteSetBstMidEndSepPunct{\mcitedefaultmidpunct}
{\mcitedefaultendpunct}{\mcitedefaultseppunct}\relax
\EndOfBibitem
\bibitem[Leon-Escamilla and Corbett(2006)Leon-Escamilla, and
  Corbett]{leon2006hydrogen}
Leon-Escamilla,~E.~A.; Corbett,~J.~D. Hydrogen in polar intermetallics. Binary
  pnictides of divalent metals with Mn$_{5}$Si$_{3}$-type structures and their
  isotypic ternary hydride solutions. \emph{Chemistry of materials}
  \textbf{2006}, \emph{18}, 4782--4792\relax
\mciteBstWouldAddEndPuncttrue
\mciteSetBstMidEndSepPunct{\mcitedefaultmidpunct}
{\mcitedefaultendpunct}{\mcitedefaultseppunct}\relax
\EndOfBibitem
\bibitem[sup()]{supplement}
See Supplemental Material for the computational methods and details for the
  constructions of the $\bm{k}\cdot\bm{p}$ effective Hamiltonian\relax
\mciteBstWouldAddEndPuncttrue
\mciteSetBstMidEndSepPunct{\mcitedefaultmidpunct}
{\mcitedefaultendpunct}{\mcitedefaultseppunct}\relax
\EndOfBibitem
\bibitem[Rauch et~al.(2017)Rauch, Minh, Henk, and Mertig]{Rauch2017Model}
Rauch,~T.; Minh,~H.~N.; Henk,~J.; Mertig,~I. Model for ferromagnetic Weyl and
  nodal line semimetals: Topological invariants, surface states, anomalous and
  spin Hall effect. \emph{Phys. Rev. B} \textbf{2017}, \emph{96}, 235103\relax
\mciteBstWouldAddEndPuncttrue
\mciteSetBstMidEndSepPunct{\mcitedefaultmidpunct}
{\mcitedefaultendpunct}{\mcitedefaultseppunct}\relax
\EndOfBibitem
\bibitem[Sun et~al.(2017)Sun, Zhang, and Chang]{Sun2017Coexistence}
Sun,~J.~P.; Zhang,~D.; Chang,~K. Coexistence of topological nodal lines, Weyl
  points, and triply degenerate points in TaS. \emph{Phys. Rev. B}
  \textbf{2017}, \emph{96}, 045121\relax
\mciteBstWouldAddEndPuncttrue
\mciteSetBstMidEndSepPunct{\mcitedefaultmidpunct}
{\mcitedefaultendpunct}{\mcitedefaultseppunct}\relax
\EndOfBibitem
\bibitem[Yang et~al.(2011)Yang, Lu, and Ran]{yang2011quantum}
Yang,~K.-Y.; Lu,~Y.-M.; Ran,~Y. Quantum Hall effects in a Weyl semimetal:
  Possible application in pyrochlore iridates. \emph{Phys. Rev. B}
  \textbf{2011}, \emph{84}, 075129\relax
\mciteBstWouldAddEndPuncttrue
\mciteSetBstMidEndSepPunct{\mcitedefaultmidpunct}
{\mcitedefaultendpunct}{\mcitedefaultseppunct}\relax
\EndOfBibitem
\end{mcitethebibliography}

\end{document}